# Wi-Fi CSI based Behavior Recognition: From Signals, Actions to Activities

Zhu Wang[1], Bin Guo[1], Zhiwen Yu[1], Xingshe Zhou[1]

wangzhu@nwpu.edu.cn

## Abstract

Human behavior recognition has been considered as a core technology that can facilitate variety of applications. However, accurate detection and recognition of human behavior is still a big challenge that attracts a lot of research efforts. Recent advances in the wireless technology (e.g., Wi-Fi Channel State Information, i.e., CSI) enable a new behavior recognition paradigm, which is able to recognize behaviors in a device-free and non-intrusive manner. In this article, we first provide an overview of the basics of Wi-Fi CSI based behavior recognition. Afterwards, we classify related applications into three-granularity: signals, actions and activities, and then provide some insights for designing new schemes. Finally, we conclude by discussing the challenges, possible solutions to these challenges and some open issues involved in CSI based behavior recognition.

## 1. Introduction

In the field of ubiquitous computing, behavior recognition is an important research topic and has been used in quite a number of human-centric services and applications, such as personalized recommendation, health monitoring, and social networking. Traditionally, to identify human behaviors, we first need to continuously collect the readings of physical sensing devices (e.g., GPS, accelerometer, and RFID), which can be either worn on human bodies, attached on objects or deployed in environments. Afterwards, based on recognition algorithms or classification models, the behavior types can be identified so as to facilitate upper layer applications. Although such traditional behavior identification approaches achieve satisfactory performances and are widely adopted, most of them are intrusive and require specific sensing devices, raising issues such as privacy and deployment cost.

With the recent advances in wireless communications, behavior recognition based on Wi-Fi has been attracting more and more attentions due to its ubiquitous availability in indoor areas. Moreover, Wi-Fi based behavior recognition approach is able to overcome the aforementioned shortcomings of traditional approaches, as it only leverages the wireless communication feature and does not need any physical sensor.

A typical Wi-Fi based behavior recognition system consists of a Wi-Fi access point (AP) and one or several Wi-Fi enabled devices in the environment. When located in indoor environment with such a system, the movement of human bodies will affect the wireless signals and change the multi-path profile of the system. Based on this principle, we are able to recognize human behaviors by exploring the changes of wireless signals caused by user movements. For example, when a person is located in the line of sight (LOS) of the Wi-Fi device and AP, the signal will be attenuated and hence a different received signal strength (RSS) is observed. Compared with RSS, the recently emerged channel state information (CSI) is a more fine-grained metric which describes both amplitude attenuation and phase shift of the wireless signal, based on which various behaviors can be recognized effectively, ranging from vital signals, basic actions to complex activities.

---

## 2. The Principle of CSI based Behavior Recognition

### 2.1. Preliminaries of CSI based Behavior Recognition

In this section, we first present the basic concepts of CSI and then give some intuition on why CSI based behavior recognition is feasible.

CSI is a metric which estimates the channel by representing the channel properties of a wireless communication link. In the frequency domain, the wireless channel can be described as $\mathbf{Y} = \mathbf{H} \times \mathbf{X} + \mathbf{N}$, where $\mathbf{X}$ and $\mathbf{Y}$ correspond to the transmitted and received signal vectors, $\mathbf{H}$ is the channel matrix presented in the format of CSI, and $\mathbf{N}$ is the additive white Gaussian noise vector.

In the IEEE 802.11n standard, CSI is measured and reported at the scale of OFDM (Orthogonal Frequency Division Modulation) subcarriers, where each $CSI_i = |CSI_i|\exp\{j(\angle CSI_i)\}$ depicts the amplitude response (i.e., $|CSI_i|$) and phase response (i.e., $\angle CSI_i$) of one subcarrier. Specifically, each entry in matrix $\mathbf{H}$ corresponds to the channel frequency response (CFR) value between a pair of antennas at a certain OFDM subcarrier frequency at a particular time, and the time-series of CFR values for a given antenna pair and OFDM subcarrier is called a CSI stream. In other words, while CFR describes the combined effects of fading, scattering and attenuation of a specific subcarrier, CSI is the union of these CFRs. Specifically, 802.11n specifications have provisions for reporting quantized CSI field per packet using various subcarrier grouping options as per clause [1, 7.3.1.27]. However, different manufacturers may choose to implement a subset of the subcarrier grouping options. For example, the Intel 5300 wireless NIC implements an OFDM system with 56 subcarriers of a 20 MHz channel or 114 subcarriers of a 40 MHz channel, 30 out of which can be read for CSI information via the device driver corresponding to 2 and 4 subcarrier grouping, respectively. Thereby, a time-series of CSI values includes $30 \times Num_{Tx} \times Num_{Rx}$ CSI streams, where $Num_{Tx}$ and $Num_{Rx}$ stand for the number of transmitting and receiving antennas, respectively.

Given an indoor environment with two wireless nodes, as shown in Fig. 1, the wireless signal will propagate in a multi-path manner, and the wireless channel will be relatively stable as long as there is no people or no motion. However, once a person moves, the scattered signals will change (the red line in Fig. 1), which causes channel disturbances, involving both amplitude attenuation and phase distortion. In other words, different multi-path effects can be obtained if a person is moving, which results in different CSI streams at the receiver and can be used to recognize different behaviors by correlating them with the corresponding channel distortion patterns.

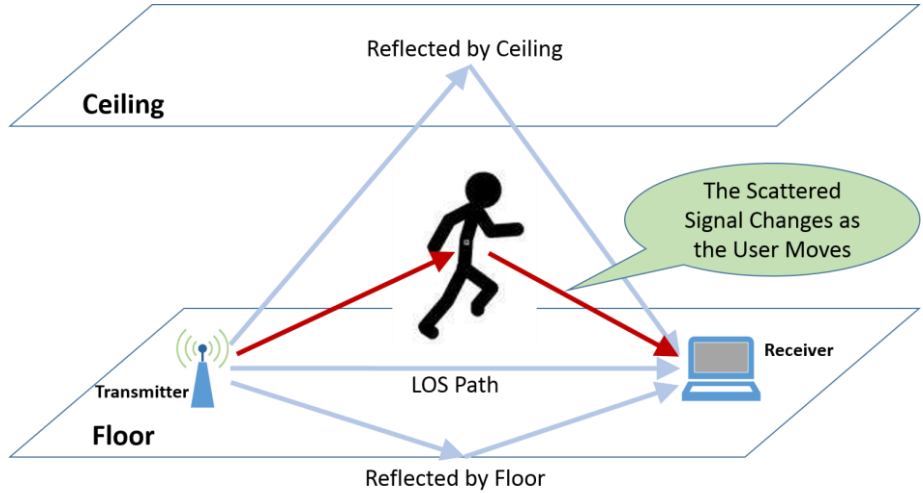

Figure 1. Wi-Fi signal propagation in indoor environments

### 2.2. Pattern-based and Model-based Behavior Recognition Approaches

Behavior recognition approaches can be categorized into two groups, i.e., pattern-based [2, 4, 6~9, 11~15] and model-based [3, 5, 10]. The pattern-based approaches aim to classify behaviors by exploring different features of CSI measurements, while the model-based approaches implement recognition by modeling the relationship between signal space and behavior space. A general architecture of Wi-Fi CSI based behavior recognition approaches is shown in Fig. 2. Though the middle part (i.e., CSI data collection and preprocessing) is common to both pattern-based and model-based approaches, the left and right parts illustrate the key difference of these two approaches.

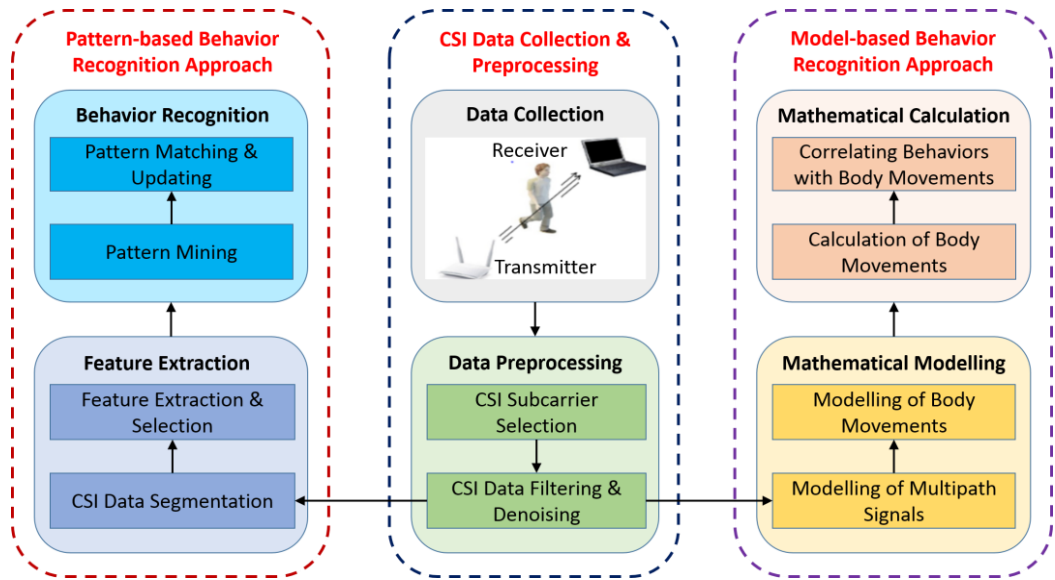

Figure 2. A general architecture of Wi-Fi CSI based behavior recognition approaches

Most existing CSI based behavior recognition studies adopt the pattern-based approach. The intuition is that different behaviors have distinct impacts on the received CSI streams, which can be leveraged to mine patterns or construct profiles for predefined behaviors, as shown in the left part of Fig. 2. Afterwards, each behavior can be classified as one of the predefined types based on profile matching or pattern recognition. The key benefit of the pattern-based recognition approaches is that they don't require intensive deployments and can work with even a single AP, which ensures the low hardware cost and maintenance and has no obstruction to human's normal life. However, pattern-based approaches usually require a learning process to construct profiles or classifiers, which restricts them to identify only a limited set of predefined behaviors.

Model-based approaches are based on the characterization of mathematical relationships between human behaviors and the received signals. In the case of Wi-Fi CSI based behavior recognition, the aim of modeling is to relate the signal space to the physical space including human and environment, and characterize the physical law through mathematical relationship between the received CSI signals and the sensing target, as shown in the right part of Fig. 2. Since the model-based approaches do not need predefined behavior profiles, they can track an arbitrary set of human behaviors, which enables wider ranges of real-time applications. Currently, there have been several model-based behavior recognition works, such as the Angle-of-Arrival (AoA) model [5], the CSI-speed model [10], and the Fresnel zone model [3].

## 3. Applications of Behavior Recognition Empowered with Wi-Fi CSI

In this section, we first classify existing studies and applications of Wi-Fi CSI based behavior recognition into three-granularity, i.e., signals, actions and activities, as shown in Fig. 3. Afterwards, based on the taxonomy and the general architecture of Wi-Fi CSI based behavior recognition approaches, we present and analyze some of the most representative studies.

According to Fig. 3, in this article, we define signals as fine-grained behaviors which mainly refer to minute and periodic movements of a certain body part, e.g., chest movements caused by heartbeat and breathing. Compared with signals, actions are medium-grained behaviors that people perform following predefined standards, e.g., gesture and sign language are actions which correspond to specific hand or finger movements with certain formations and directions. Differently, activities refer to coarse-grained complex behaviors that neither follow any periodicity nor standard, such as in-place activities (e.g., cooking dinner and exercising on a treadmill) and walking related activities (e.g., running and falling). Thereby, we can conclude that human behaviors become more complex and irregular as the granularity varying from fine to coarse. For instance, an activity such as cooking usually consists of a set of movements to fetch, prepare, and mix ingredients that may occur in different sequences, making it hand to recognize with identification methods that are proposed for single movements or gestures [9].

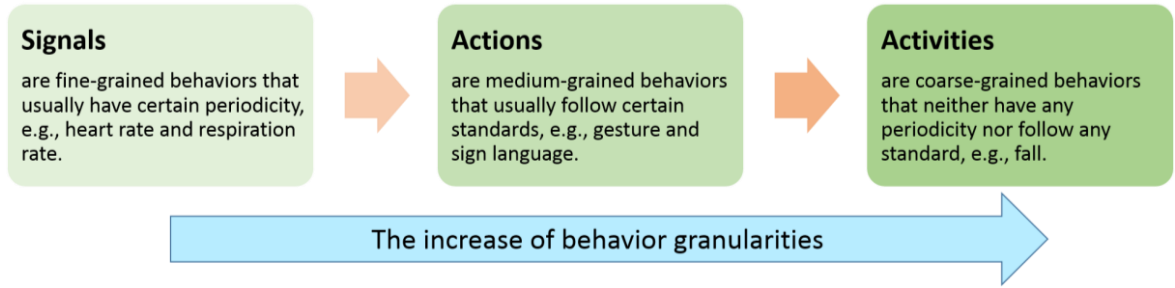

Figure 3. A taxonomy of human behaviors

### 3.1. Signal Recognition

In this article, signals mainly refer to vital information such as respiration rate and heart rate. A summary of studies on the recognition of vital signals is given in Table 2.

Table 1. A summary of studies on signal recognition.

| | Behaviors | Number of Users | Data Preprocessing Methods | Features | Recognition Approach | Test Bed and Data Set | Performance |
|---|---|---|---|---|---|---|---|
| Liu et al. [2] | Respiration. | Single. | Hampel identifier, linear interpolation, and wavelet filter. | Amplitude as well as the periodicity level of different CSI sequences to extract the rhythmic patterns associated with respiration. | Pattern-based approach using short-time Fourier transform. | The system is evaluated in a typical office. | The detection rate is greater than 85% for all 6 sleeping positions. |
| Zhang et al. [3] | Respiration. | Single and two. | Hampel filter and moving average filter. | Chest displacements are converted to phase changes, based on which the time-varying amplitude of resultant receiving signals (a sinusoidal wave) is used to characterize human respiration. | Model-based approach using the Fresnel zone model. | 9 participants over three months. | While the user is located in the middle of a Fresnel zone ellipse, the accuracy can reach 100%. |
| Liu et al. [4] | Breathing rate and heart rate. | Single and two. | Hampel filter and moving average filter. | Time domain and frequency domain features to characterize periodic minute movements at different frequency domains. | Pattern-based approach using $k$-means. | 6 participants over three months. | In case of the breathing rate, 80% estimation errors are lower than 0.5bpm; in case of the heart rate, 90% of estimation errors are less than 4bpm. |

*Respiration rate.* As CSI provides fine-grained information, it can be leveraged to recognize vital signals such as respiration rate and heart rate. Based on off-the-shelf Wi-Fi devices, the Wi-Sleep [2] system is able to extract a user's respiration information under various sleeping positions, by identifying the rhythmic patterns associated with respiration. Based on the Wi-Fi Fresnel zone model, Zhang et al. [3] developed a theory that is able to explain the detectability of respiration from the aspects of breathing depth, location and orientation. With the proposed theory, not only when and why Wi-Fi CSI can be used to detect respiration become clear, it also reveals the physical limit and foundation of such wireless sensing systems.

*Heart rate.* Compared with respiration rate, heart rate is a more minute movement with higher frequency, making it more difficult to recognize. Liu et al. [4] developed a system to simultaneously track both heart rate and breathing rate by exploring CSI streams. In particular, the proposed system is able to capture vital signs when either one or two users are in bed. Experiments under realistic settings show that the system can accurately detect both heart rate and breathing rate during sleep, and achieve comparable performance comparing with existing approaches.

### 3.2. Action Recognition

In this article, actions mainly refer to human behaviors that can be used for interaction, such as gestures, talks, keystrokes, etc. A summary of studies on the recognition of such actions is given in Table 2.

Table 2. A summary of studies on action recognition.

| | Behaviors | Number of Users | Data Preprocessing Methods | Features | Recognition Approach | Test Bed and Data Set | Performance |
|---|---|---|---|---|---|---|---|
| Sun et al. [5] | Hand motions, e.g., drawing arbitrary lines and curves. | Single. | Threshold based filtering. | The azimuth and elevation of AoAs are used to determine the horizontal and vertical coordinates as well as their depth along the hand's trajectory. | Model-based approach using the AoA model. | 90,000 trajectories from 10 users. | The median error for tracking user hand is lower than 5 cm, and the average recognition accuracy are 95% and 91% for letters and words. |
| Li et al. [6] | Finger gestures, e.g., digits 1-9 in ASL. | Single. | Hampel identifier and Butterworth low-pass filter. | A compressed synthetic waveform (i.e., gesture profile) is defined as the feature vector. | Pattern-based approach using the dynamic time warping method and the KNN classifier. | 10 users. | The average classification accuracy for 9 digits in ASL is 90.4%. |
| Wang et al. [7] | Human talks (i.e., lip reading and speech recognition). | Multiple. | Butterworth IIR band-pass filtering. | Based on partial multipath effect, Wavelet-based mouth motion profiles are extracted as features to characterize mouth moving patterns. | Pattern-based approach using the dynamic time warping method. | 4 participants in 6 scenarios. | In case of one user speaking no more than 6 words, the accuracy is up to 91%; in case of no more than 3 users talking simultaneously, the accuracy is about 74%. |
| Ali et al. [8] | Keystroke. | Single. | Low pass Butterworth filter. | High resolution CSI-waveform shapes are used to characterize the unique multi-path distortions of different keystrokes. | Pattern-based approach using the dynamic time warping method and the KNN classifier. | 10 users. | The detection rate for the keystroke is 97.5% and the classification accuracy for single keys is 96.4%. |

*Human gestures*. A number of studies have investigated how to recognize human gestures with Wi-Fi CSI, which can be used as a convenient human-computer interaction mode. WiDraw [5] is a hand motion tracking system that leverages the Wi-Fi signal's AoA values at the mobile devices to trace hand trajectories. The intuition of WiDraw is that whenever a signal from a specific direction is affected by the user's hand, the signal strength of the angle denoting the same direction will decline. Similarly, the Wi-Finger [6] system aims to recognize finger gestures (e.g., digits 1-9 in ASL) using ubiquitous wireless signals. The approach is based on the observation that a user's fingers move in a particular formation and direction while performing a certain gesture, leading to a unique pattern in CSI streams.

*Human talks*. Compared with human gestures, human talk based interactions would cause even less burden to the user. Wang et al. [7] developed WiHear by exploring Wi-Fi CSI to "hear" human talks. In particular, to capture reflections caused by mouth movements, the authors introduced Mouth Motion Profile that leverages partial multipath effects and wavelet packet transformation. Specifically, WiHear can recognize talks within the range of Wi-Fi signals, as such signals do not require LOS path. Furthermore, WiHear is able to "hear" multiple users simultaneously based on the MIMO technology.

*Keystrokes*. The intuition of WiKey [8] is similar to that of Wi-Finger [6], i.e., a person's hands and fingers move in a specific formation and direction while typing a certain key, generating a unique pattern in CSI streams. In particular, when a user types on a keyboard, the typed keys are identified by recognizing the way how CSI values change. According to experimental results, WiKey achieves more than 97.5% and 96.4% accuracy for keystroke detection and single key classification, respectively.

## 3.3. Activity Recognition

In this article, activities refer to human behaviors such as walking, fall, human identity, etc. A summary of daily activity recognition related studies and applications is given in Table 3.

Table 3. A summary of studies on daily activity recognition.

| | Behaviors | Number of Users | Data Preprocessing Methods | Features | Recognition Approach | Test Bed and Data Set | Performance |
|---|---|---|---|---|---|---|---|
| Wang et al. [9] | In-place activities and walking movements. | Single. | The dynamic exponential smoothing filter. | Time and frequency domain features of the amplitude (e.g., distributions of the CSI measurements). | Pattern-based approach using the dynamic time warping method and the earth mover distance technique. | 4 volunteers performed 9 typical in-place activities and 8 walking activities. | The average true positive rate is 96% when using 3 devices, and the detection rate is around 92% when with only one device. |
| Wang et al. [10] | Human activity. | Single. | PCA based CSI denoising scheme. | A 27 dimensional feature vector in the frequency domain. | Model-based approach using the CSI-speed model. | 25 volunteers performed 8 different activities. | The average accuracy is 96.5% for trained places and trained users, and about 80% for places and persons that have not been trained on. |
| Wang et al. [11] | Gait patterns and human identity. | Single. | PCA. | A set of gait features, including walking speed, gait cycle time, spectrogram signatures, et al. | Pattern-based approach using LibSVM. | 2,800 gait instances from 50 subjects. | Over 50 subjects, the recognition accuracies are 79.28%, 89.52%, and 93.05% for top-1, top-2, and top-3 candidates. |
| Zeng et al. [12] | Gait patterns and human identity. | Single. | Butterworth bandpass filter. | Time domain and frequency domain features are extracted to characterize a person's gait. | Pattern-based approach using decision tree. | 20 volunteers at 3 different locations. | The recognition accuracy is 92% to 80% for 2 to 6 human subjects. |
| Wang et al. [13] | Fall. | Single. | 1-D linear interpolation algorithm and band-pass filter. | A set of 8 features are extracted from both CSI amplitude and phase difference. | Pattern-based approach using SVM. | 6 volunteers over two months. | The sensitivity and specificity are 91% and 92%. |
| Zhang et al. [14] | Human identity. | Single. | A Butterworth filter and a silence removal method. | Time domain and frequency domain features are extracted to characterize a person's walking style. | Pattern-based approach using sparse approximation classification. | 10 subjects for training and another 20 subjects for testing. | The accuracy of human identification is 93% to 77% for 2 to 6 individuals. |
| Xin et al. [15] | Human identity. | Single. | Butterworth IIR filter. | Approximation coefficients are used to represent shape features of the LOS waveform. | Pattern-based approach using the dynamic time warping method and the KNN classifier. | 9 volunteers, and each of them provided 40 samples. | The accuracy of human identification is 94.5% to 88.9% for 2 to 6 individuals. |

*Ordinary daily behaviors.* Based on Wi-Fi access points and devices, Wang et al. [9] developed a device-free activity recognition system named E-eyes, which is capable of identifying both in-place activities as well as walking movements by comparing them against predefined profiles. Similarly, with the proposed CSI-speed model and CSI-activity model, the CARM system [10] first extracts the relationship between CSI dynamics and user activities and then utilizes such correlation to match each activity to the best-fit profile.

*Walking related behaviors.* Typical walking related behaviors that can be recognized based on CSI include walking, falling, etc. For example, Wang et al. [11] developed a gait pattern recognition system, which can extract fine-grained gait information, including walking speed, footstep length, gait cycle time, etc. Similarly, Zeng et al. [12] also used Wi-Fi CSI to identify a person's steps and walking gait. Another walking related behavior that attracts the attention of quite a number of researchers is fall. For example, RT-Fall [13] is a representative CSI based fall detection system, which explores both the amplitude and phase of CSI measurements.

*Human identity.* Zhang et al. [14] developed WiFi-ID, a device-free system that uses off-the-shelf devices to identify individuals. The basic idea is that each individual has a unique walking style and body shape which causes unique disturbances in the Wi-Fi signals and can be characterized with the features extracted from CSI. There are several similar studies, e.g., the FreeSense [15] system achieves indoor human identification by comparing shape features of the LOS waveform, and the WiWho [12] system identifies a person from a small group of people (2 to 6) based on gait analysis.

### 3.4. Insights for Designing New Schemes

According to the above classification and summarization, we can obtain the following insights:

*CSI features for different behaviors.* Different features might be suitable for recognizing behaviors of different granularities. To recognize fine-grained behaviors (i.e., signals) that have certain periodicity, data calibration and subcarrier selection are needed to select subcarriers that are sensitive to minute body movements, so as to extract periodic or rhythmic features of the CSI data. When performing actions, a certain body part (e.g., hands or fingers) would move in a specific formation and direction (based on the predefined standards) and thus result in a unique pattern in CSI steams. Thereby, we need to extract sequence features to distinguish different actions. Particularly, to achieve satisfactory recognition performance without affecting the quality of communication, we can adopt certain techniques (e.g., band-pass filtering and correlation analysis) on the receiving device to reduce the data caused by irrelevant multipath effects. Compared with signals and actions, coarse-grained behaviors usually do not follow explicit periodicity or patterns. For example, wireless signals reflected from different human body parts have distinct frequencies, as the moving speed of different body parts varies. Thus, to effectively distinguish different complex behaviors, we need to extract more sophisticated features to characterize the movements of different body parts.

*Recognition approaches for different behaviors.* According to Table 1~3, we can find that while the pattern-based approach can be used to recognize behaviors of all the three-granularity, it is still a challenge to design model-based approaches, especially for complex behaviors. Specifically, a typical fine-grained behavior usually is related to tiny and periodic movements of a certain body part (e.g., chest movement caused by respiration). Therefore, its disturbance in the CSI data is relatively stable and clear, and it is much easier to theoretically characterize the correlation between such a behavior and CSI dynamics. For example, the Fresnel model can perfectly explain when and why human respiration is detectable [3]. On the contrary, due to the simultaneous and irregular movements of different body parts, most coarse-grained behaviors would cause complex disturbances in the CSI data, making it difficult to build mathematical models for such behaviors, which might be the reason why most existing studies on the recognition of complex activities adopt pattern-based approaches.

*Recognition performances of different behaviors.* Generally speaking, as shown in the last column of Table 1~3, with the increasing of behavior complexity (i.e., from signals, actions to activities), the recognition performance would decline gradually. For instance, the detection accuracy of respiration rate (a representative signal) can reach 100% based on the Fresnel model [3], the average classification accuracy for 9 digits in ASL (a representative action) is above 90% [6], while the identifying accuracy of 6 individuals (a representative activity) is around 80% [12, 14]. Moreover, compared with signals and actions, the recognition of complex activities is more sensitive to the influence of environment and user changes. For example, while a respiration rate detection method has stable performance for different users [3], the performance of a gait pattern recognition method would differ significantly [11].

## 4. Limitations, Challenges, and Open Issues

Based on the classification and analyzing of existing studies on Wi-Fi CSI based behavior recognition as well as the obtained insights, we identify the limitations, challenges and open issues as follows.

### 4.1. Theoretical Foundation for CSI-based Behavior Recognition

While most existing CSI-based behavior recognition systems adopted the pattern-based approach, a key limitation of such systems is the lack of a theoretical model that is capable of quantitatively correlating CSI dynamics and user behaviors. To distinguish different behaviors, these systems usually construct a set of profiles using the statistical characteristics of Wi-Fi wireless signals, such as time domain and frequency domain features of CSI streams. Meanwhile, as the changes of CSI values caused by human behaviors are not always linear, non-linear and non-stationary features (e.g., the sample entropy and the coefficient of detrended fluctuation analysis) are also important for the characterization of behavior patterns. However, such an approach relies on the assumption that the correspondence between profiles and behaviors is unique, which cannot be proved theoretically. Therefore, in order to develop widely acceptable systems, one of the most important challenge is revealing the principle reason of signal changes due to body movements, and mathematically modeling the relationship between CSI dynamics and user behaviors. Without such a model, it is hard to optimize the performance of CSI-based behavior recognition systems. Recently, there has been encouraging progress in addressing this issue. For example, the Wi-Fi Fresnel zone model [3] can be used to explain when and why human respiration is detectable based on Wi-Fi CSI. Nevertheless, more efforts are still needed to fully understand the theoretical limit and foundation of CSI based behavior recognition systems, especially systems for complex and irregular behaviors. One possible solution is to combine the advantages of existing models (e.g., the Angle-of-Arrival model [5], the CSI-speed model [10], the Fresnel zone model [3]), so as to construct a more generic and powerful method.

### 4.2. Behavior Recognition with Individual Differences

According to Table 1~3, we observe that the experiment of most CSI-based behavior recognition systems is based on quite limited amount of participants, which is another limitation of these systems as the recognition performance may severely decline for untrained users. For example, the average recognition accuracy of the CARM [10] system declines from 96.5% to 80% as the target users change from trained ones to untrained ones. The reason is that different individuals usually have distinct body characteristics or behavior habits, leading to different impact to the received CSI even when performing the same behavior. Thereby, another challenge for CSI-based behavior recognition is how to overcome individual differences and build universal behavior profiles or models. One possible approach to address this issue is to leverage techniques such as deep neural network to extract more efficient features.

### 4.3. Effective Recognition of Complex Activities

As aforementioned, compared with vital signals or actions that either have certain periodicity or follow certain standard, complex activities are more difficult to recognize accurately, because different activities might have similar profiles. In other words, the recognition of such similar activities might correspond to an underdetermined system of equations, i.e., an ill-posed problem. For example, features extracted for fall-like activities (e.g., quickly sits down) can be quite similar to those of falls under certain scenarios [13], leading to incorrect recognition results. Therefore, the accurate recognition of complex activities is another challenging issue. To improve the system's performance, a promising approach is to develop novel recognition algorithms by introducing contextual information of the physical environment as a form of constraint to regularize an otherwise ill-conditioned model. For instance, in case of fall detection, we can first obtain the user's location, and then adopt a location-based classification model to achieve effective recognition.

### 4.4. System Flexibility

Currently, most CSI-based behavior recognition applications are still at the lab prototype stage, and one key limitation is the flexibility of these systems. Specifically, due to the multi-path effects of the environment and users, the performance of a CSI-based behavior recognition system is sensitive to hardware deployment (e.g., antenna locations and the number of Wi-Fi APs), changes in the environment (e.g., the location of furniture) and the presence of other users or pets. For example, among the 14 systems summarized in Table 1~3, only 3 of them [3, 4, 7] can simultaneously recognize behaviors of two or multiple users and none of them can support simultaneous recognition of complex activities. Thereby, the simultaneous recognition of multiple human's behaviors is still an open issue, especially the recognition of complex activities. One possible approach is to build adaptive behavior profiles or models for different environments.

### 5. Conclusion

In this article, after a briefly introduction to the basics of Wi-Fi CSI based behavior recognition, we propose to classify and analyze CSI-based behavior recognition studies and applications from three-granularity, i.e., signals, actions and activities. In particular, signals are fine-grained behaviors which refer to minute and periodic body movements, e.g., respiration rate and heart rate. Actions are medium-grained behaviors that people perform following certain standard, e.g., gesture and sign language. Activities refer to coarse-grained behaviors that neither follow any periodicity nor standard, e.g., fall. Based on this taxonomy, we provide some insights for designing new schemes, and further identify and discuss the challenges and open issues of CSI based behavior recognition.

## Acknowledgments

This work was supported in part by the National Key R&D Program of China (No. 2016YFB1001401), the National Natural Science Foundation of China (No. 61332013, 61402369), and the Natural Science Foundation of Shaanxi Province (No. 2015JQ6237). The authors would like to thank the anonymous reviewers for their valuable comments.

## Biographies

**ZHU WANG** [M] (wangzhu@nwpu.edu.cn) is an associate professor of computer science at Northwestern Polytechnical University, China. He received his B.Eng, M.Eng and Ph.D. degree of Engineering in computer science and technology in 2006, 2009 and 2013 respectively from the same university. During 2010-2012, he was a research fellow at Institut TELECOM SudParis in France. His research interests include pervasive computing, social network analysis, and health informatics.

**BIN GUO** [M] (guob@nwpu.edu.cn) is a full professor of computer science at Northwestern Polytechnical University, China. During 2009-2011, he was a post-doctoral researcher at Institut TELECOM SudParis in France. His research interests include pervasive computing, social computing, and mobile crowd sensing. He has served as an editor or guest editor for a number of international journals, such as IEEE Communications Magazine, IEEE THMS and IEEE IT Professional.

**ZHIWEN YU** [SM] (zhiweny@nwpu.edu.cn) is a full professor of computer science at Northwestern Polytechnical University, China. He has worked as an Alexander Von Humboldt Fellow at Mannheim University, Germany from Nov. 2009 to Oct. 2010, and a research fellow at Kyoto University, Japan from Feb. 2007 to Jan. 2009. His research interests cover pervasive computing, context-aware systems, and personalization. Dr. Yu has served as an editor or guest editor for a number of journals, such as IEEE Communication Magazine, IEEE THMS and ACM TIST.

**XINGSHE ZHOU** [SM] (zhouxs@nwpu.edu.cn) is a full professor of computer science at Northwestern Polytechnical University, China. During 1995-2011, he served as the dean of the school of computer science at Northwestern Polytechnical University. His research interests include cyber-physical system, pervasive computing, and cloud computing.

**Contact address (Zhu Wang)**: Mailbox #404, Northwestern Polytechnical University, No. 127, West Youyi Road, Xi'an City, Shaanxi Province, 710072, China